# Optimizing noise level for perturbing geo-location data

Abhinav Palia
palia@usc.edu
University of Southern California

Rajat Tandon
rajattan@usc.edu
University of Southern California

**Abstract**

With the tremendous increase in the number of smart phones, app stores have been overwhelmed with applications requiring geo-location access to provide their users better services through personalization. Revealing a user's location to these third-party apps, no matter at what frequency, is a severe privacy breach which can have unpleasant social consequences. In order to prevent inference attacks derived from geo-location data, a number of location obfuscation techniques have been proposed in the literature. However, none of them provides any *objective* measure of privacy guarantee. Some work has been done to define differential privacy for geo-location data in the form of geo-indistinguishability with *l* privacy guarantee. These techniques do not utilize any prior background information about the Points of Interest (*PoI*s) of a user and apply Laplacian noise to perturb all the location coordinates. Intuitively, the utility of such a mechanism can be improved if the noise distribution is derived after considering some prior information about *PoI*s.

In this paper, we apply the standard definition of differential privacy on geo-location data. We use first principles to model various privacy and utility constraints, prior background information available about the *PoI* s (distribu--tion of PoI locations in a 1D plane) and the granularity of the input required by different types of apps, to produce a more accurate and a utility maximizing differentially private algorithm for geo-location data at the OS level. We investigate this for a category of apps and for some specific scenarios. This will also help us to verify that whether Laplacian noise is still the optimal perturbation when we have such prior information.

## 1 Introduction

Over the years, several mobile phone services are becoming dependent on user's location to provide a better experience, be it a dating app, restaurant search, nearby gas stations lookup and what not. All these services require a user to surrender her location (mostly exact coordinates) to derive accurate results. With the increasing popularity of social networks, extracting auxiliary information about an individual has become easier than ever before. Both factors have increased the likelihood of inference attacks on the users, which can have unpleasant social consequences. Therefore, revealing user's location, no matter at what frequency, is a severe privacy breach.

The criticality of geo-location data can be estimated by the news pieces reporting that the Egyptian government used to locate and imprison users of Grindr–a gay dating app [4]. Grindr uses geo-location of its users to provide them a perfect match in their vicinity. Most of the users have submitted their stats such as body weight, height, eye color, ethnicity, preferences, on-prep (AIDS status), extra information etc. while creating a profile. Even half of these values along with their geo-location, can be used to derive inferences uniquely identifying a user.

[6] has reported social relationship leakage of a user through applications which use GPS data. Several inferences can be deduced by observing social relationships of an individual which he might not want to disclose.

Tracking location coordinates or identifying *PoI* s of an individual, can characterize his mobility and can infer sensible information such as hobbies, political, religious interests or even potential diseases [7]. All these studies provide enough motivation for the research community to find a solution to protect geo-location privacy.

Although geo-indistinguishability presents various appealing aspects, it has the problem of treating space in a uniform way, imposing the addition of Laplace noise everywhere on the map [3]. This assumption is too strict and can affect the utility of the service. A Laplace-based obfuscation mechanism satisfying this privacy notion works well in the case where no prior information is available. However, most of the apps which require geo-location as input, are conditioned with the prior of the destination or the *PoI* s, in general.

In this paper, we would try to investigate that whether the choice of using Laplacian noise to perturb geo-data is optimal in the scenarios where prior information about user's PoIs is available. Intuitively, availability of this information will improve the utility of the differential private mechanism but must be conditioned with some more constraints. In the next section, we discuss the related work done in this direction. In section 3, we clearly define the problem statement we counter in this paper. Section 4 discusses our proposal and the contribution towards the solution of this problem. In section 5 and 6, we present our results and the future trajectory of our work, respectively. In section 7, we conclude our findings. After listing the references, appendix A and B provide the mathematical solution of our constraints.

## 2   Related Work

Most of the hand-held devices provide three options of allowing location access to the installed apps, namely– *Always, While using* and *Never*. One can easily predict the harm which can be caused when this permission is granted Always. On the other hand, we still want to use the service from the app, so not allowing this permission by selecting *Never* is not a valid choice. In such a situation, *While using option* appears appropriate but can still be used by an attacker to track the trajectory of a user. Intuitively, it is better to trust the OS which can sanitize the geo-location data before supplying it as an input to the app.

Literature proposes different ideas to perturb geo-location data. [10] proposes the idea of spatial and temporal cloaking which uses *k* –anonymity, *l* –diversity and *p*–sensitivity. Other spatial obfuscation mechanisms proposed in [11, 12] reduce the precision of the location infor- mation before supplying it to the service. Most of these techniques are not robust and are also detrimental to utility functions [1] as they are based on very simple heuristics such as i.i.d. location sampling or sampling locations from a random walk on a grid or between points of in- terest. The generated location traces using these types of techniques fail to capture the essential semantic and even some basic geographic features. Techniques such as spatial cloaking perturb the exact location of the user but do not provide any privacy guarantee. Additionally, they are not resistant to probability based inference attacks [9].

Thus, there exists some knowledge gap between these techniques and the desired characteristics of a location perturbation mechanism. Differential privacy holds a good reputation in providing a privacy guarantee by adding care- fully calibrated noise which also maintains an acceptable level of utility. Geo-indistinguishability proposed in [2], defines a formal notion of privacy for location-based systems that protect user's exact location, while allowing approximate information–typically needed to obtain a certain desired service to be released. It formalizes the intuitive notion of protecting the user's location within a radius *r* with a level of privacy *l* that depends on *r*, and corresponds to a generalized version of the well-known concept of differential privacy. The authors in [2] claim that adding Laplace noise, can perturb data effectively. As pointed out in [8], the utility of a differentially private mechanism can be increased if some prior information is available about the user. Also in [13], a *generic prior* distribution π, derived from a large user dataset is used to construct an efficient remap function for increasing the utility of the obfuscation algorithm. Clearly, if we can gather some information about *PoI* s of a user, it can help us to provide a more useful result. However, this information leakage (prior distribution available publicly) is useful for the adversary to design his remap function over the output of a differentially private mechanism. Therefore, the privacy bounds would require some alteration and intuitively, use of Laplace noise might not be an optimal choice. Now in the next section, we state our problem statement.

## 3  Problem Statement

In this section, we clearly define the problem statement. Since geo indistinguishability is a flavor of differential privacy for geo location data, it does not take into account various factors such as (i) *π*: denotes the priori probability distribution (prior), which is relative to the user and her knowledge (based on user's PoI history) [8]; (ii) *ψ*: denotes the priori probability distribution relative to OS's knowledge about the location of the *PoI* s, for instance, location of restaurants (= *PoI* ) relative to the current location. Since most of the LBS require the user to provide the " *destination location*", (through which OS can determine *ψ*), this information can help the OS to perturb the original location in a biased way (towards the *PoI* ) and therefore maximize the utility of the mechanism. Clearly, with the knowledge about ψ, we can have a set of linear constraints and can use them to determine whether Laplace is still the best choice or do we need to have some new noise distribution.

In the next sections, we will begin by stating the basics and then will use first principles to model various privacy and utility constraints.

## 4  Our Proposal

First we define the basic structure of the problem by stating the prior, privacy and utility goals of the mechanism. Then we define the example problem and state the privacy–utility constraints.

**PoI prior ($\psi$):** For multiple $PoIs$ located at $L_1, L_2, ...$ from the actual location $i$, the prior is defined as the distribution of these $PoIs$, denoted by $\psi = \{L_1, L_2, ...\}$

**Privacy**: We use [2] to define the notion of privacy, *i.e.*, for any user located at point $i$, she enjoys $\rho$-*privacy* within a radius $r$. More precisely, by observing $z$, the output of the mechanism $K$ when applied to $i$ (as compared to the case when $z$ is not available), does not increase the attacker's ability to differentiate between $i$ and $j$ ($|i-j| \leq \delta$ and $j$ lies inside the circle of radius $r$ centered at $i$) by more than a factor depending on $\rho$ ($\rho = \epsilon.r$).

**Utility**: We propose that a differentially private mechanism for geo-location data is utility maximizing if the output of a Location Based Service (LBS) does not change if the input given to

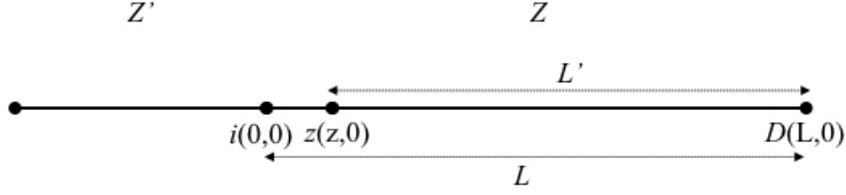

Figure 1: 1-Dimensional scenario

it is the perturbed location (as compared to the output when the input is the original location). The output of a LBS depends on the type of query it answers and the scope of these queries is vast. However, in this paper we restrict ourselves to 2 queries-

**Query 1**: *Get me the nearest PoI, my distance to it and provide the option to navigate to it.*

The output should be the nearest PoI and the approximate/effective distance to it when a perturbed input location is provided to the LBS

**Query 2**: *Get me the nearest PoI and my distance to it.*

The output should be the nearest PoI and the approximate/effective distance to it when a perturbed input location is provided to the LBS

**Query 3**: *Get me the list of PoIs starting from the nearest to the farthest.*

The output of this query is to provide the list of the PoIs, from the nearest to the farthest, such that the order of the output list is same when a perturbed location input is provided or when exact location was provided.

**Example Problem**

For the sake of simplicity, we begin with a 1- Dimensional example problem (Figure 1) in which a user Alice is located at a point $i$ and the point of interest is the restaurant located at $L$ distance from her. Further, we shift the origin at $i$, so we can denote the coordinates of the destination as $(L, 0)$ and we can write $\psi = \{L\}$. Alice wants to have $\rho$ level of privacy within a distance $r$ from $(0, 0)$. We define privacy level $\rho$ within a linear distance $r$ on both sides of the original location instead of a circle with radius $r$ just to suit our 1 D model. We also want to ensure $\rho_0$ level of privacy outside this region. For differential privacy to hold, we consider a mechanism $K$, conditioned with $\psi$, which takes location $i$ as input and produces output $z$ from the output space $S \subseteq \mathbb{E}$, ($\mathbb{E}$ is 1-D Euclidean plane). $S$ in this case includes all the points lying on the $x$-axis. Intuitively, availability of a prior $\psi$, will help us to provide a better output but will not affect the privacy constraints.

**Privacy Constraints**: The *privacy* constraints for our mechanism (both queries) are as follows:

(i) $P(i, z, \psi) > 0; \forall z$;

   The probability of outputting any points in the output space when the mechanism is applied on the input location $i$ should be non zero

(ii) $\sum_{z=-\infty}^{\infty} P(i, z, \psi) = 1$;

   Probability values must sum to 1, given that for our case $z \in S \subseteq \mathbb{E}$. $S$ in this case includes all the points lying on the $x$-axis

(iii) $P(i, K(i) \in S, \psi) \leq e^{\rho}.P(j, K(j) \in S, \psi)$, for two points $i$ and $j$ with $|i-j| \leq \delta$;

   Differential Privacy constraint derived from the definition [2], where $\epsilon$ is related to user defined level of privacy $\rho$ as $\rho = \epsilon.r$

**Utility Constraints**

1. *Query 1 and 2*: For better understanding of the solution, we categorize the apps into two classes–*Class A*: Apps which output the nearest restaurant, the distance to it and also have the feature of providing navigation to this *PoI*. *Class B*: Apps which output the nearest restaurant and the distance to it but do not provide navigation facility (and do not show the original location of the user). The logic behind having these two classes is that the utility constraints are more relaxed in the *Class A* apps since the output space is less restricted and we can have the same distance to the *PoI* from a number of places. While in case of *Class B* apps, the app should have to be supplied with a point more near to the original location because it will need a starting point for the navigation to start. Summarizing,
   *Query 1*– Apps which will take perturbed user location as input and output the nearest PoI, the distance to it and will provide them the option to start navigation (*Class B*).
   *Query 2*–Apps which will take perturbed user location as input and output the nearest PoI and the distance to it (no navigation) (*Class A*).

   - *Query 1*:
     (i) $P(i, z, \psi) > P(i, -z, \psi)$; where $-z$ denotes the points in the opposite direction of the prior
     (ii) Minimize the distance between the output point and the original location; For maximum utility
   - *Query 2*:
     (a) $i \leq z \leq L$
        (i) $P(i, z, \psi) > P(i, z_1, \psi); z < z_1$
        (ii) $Minimize ||L - i| - |L - z||$
     (b) $L \leq z \leq 2L$
        (i) $P(i, z, \psi) > P(i, z_1, \psi); z > z_1$
        (ii) $Minimize ||L - i| - |z - L||$
     (c) $-\infty \leq z \leq i$
        (i) $P(i, z, \psi) > P(i, z_1, \psi); z > z_1$
        (ii) $Minimize ||L - i| - |L - z||$
     (d) $2L \leq z \leq \infty$
        (i) $P(i, z, \psi) > P(i, z_1, \psi); z < z_1$
        (ii) $Minimize ||L - i| - |z - L||$

   We solve these constraints in the Appendix A and B.

2. *Query 3*: In these queries, the app outputs the list of the PoIs, from the nearest to the farthest, such that the order of the output list is same when a perturbed location input is provided and when exact location was provided. Here we define tolerance limits $m$, which is the distance from the starting point such that if we output point $z$ within this space, the ordering of the output remains intact. Since the *PoI*s are at different distances in the two directions ($+x-axis, -x-axis$), there will be two different tolerance limits–for the $+x$ ($m_+$) and for $-x$ ($m_-$). The probability of outputting $z$ should be maximum within this region satisfying the $Minimize.|z-i|$ constraint and after $m_+$ and $m_-$, there will be steep decline (not zero) because it will kill the utility (making it 0 will decrease privacy).

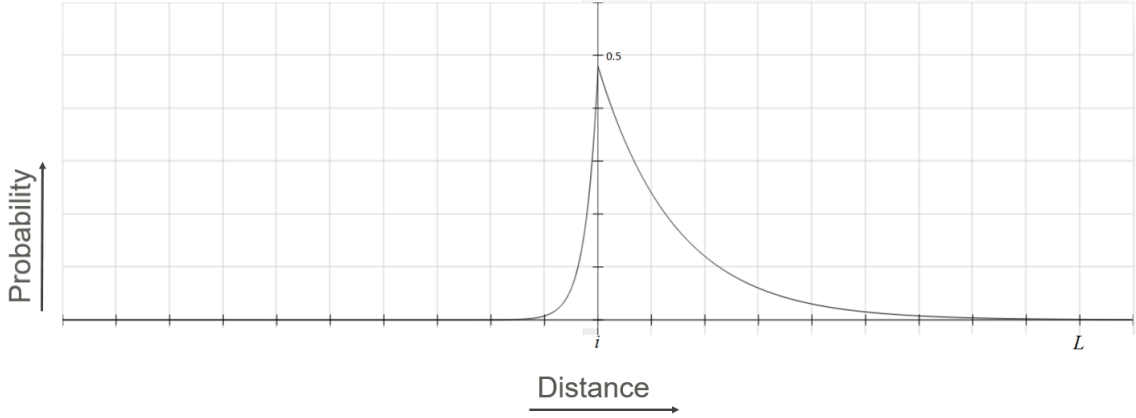

Figure 2: Probability distribution of output points for query 1

Based on above, the utility constraints are as follows:

(i) $Minimize \ |z-i|$
(ii) $P(i, m_- \leq z \leq m_+, \psi) > P(i, z > m_+, z < m_-, \psi)$

However, there can be multiple factors which can govern the outcome for this case. For instance, the output location should be in the direction where there are more *PoI*s or closeness of certain *PoI*s to the original location. Taking these factors into consideration, the complexity of this case increases and therefore, we reserve the solution of the constraints for *query 3* as our future work.

## 5 Results

In this section, we draw graphs for probability distribution for the output points for query 1 and query 2 which can be used to add noise to the original location. We use $\rho = ln2$, which implies that user wants $\rho$ level of privacy within some distance $r$ and as described earlier, $\rho = \epsilon.r$. Based on the derivation in Appendix A and B, we have the maximum probability value $p \leq 0.48$ for this case and use the approximation parameter $\alpha = 4$. Using these values, we have Figure 2 for *query 1* and Figure 3 for *query 2*. As predicted, the curve in Figure 3 is symmetrical about the destination prior at point $(L, 0)$.

## 6 Future Work

As our future work, we plan to take into account multiple factors which can be used to define the output solution for multiple *PoI*s (*query 3*). Further, we would want to extrapolate this work for 2 dimensions and then for 3 dimensions so that it is applicable to the real location data. We plan to make a google chrome extension using our 3D mechanism to test it further before it can be deployed for different kinds of mobile apps.

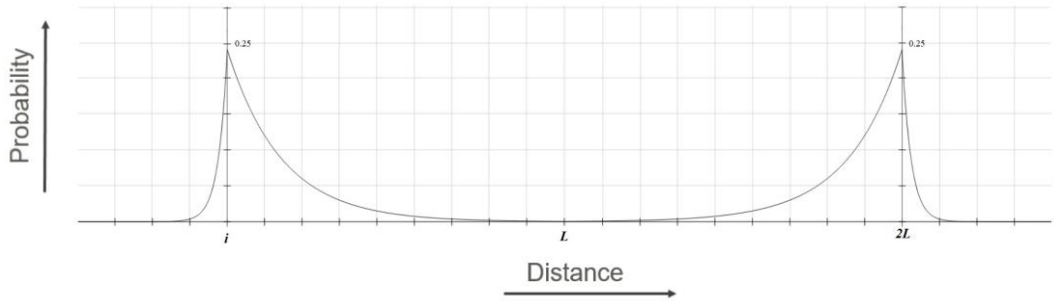

Figure 3: Probability distribution of output points for query 2

# 7 Conclusion

In this paper, we have worked on improving the utility of a differentially private mechanism for geo-location data. We have used the notion of geo-indistinguishability to provide differential privacy guarantee for geo-location data and at the same time, we have used the prior information available to the OS about the *PoI* s in order to improve the utility of the mechanism. Through mathematical formulation of the problem and solving the linear system of constraints, we have derived the probability distribution of the output points, which can be used to add noise to the original input location accordingly. Through our results, it is clear that Laplace is not the optimal choice for geo location queries conditioned with a prior and with our mechanism we have strived for maximum utility for 2 queries. We have further discussed our future work which lays the trajectory of what we plan to do next in order to have an optimum solution for the real world geo location data. According to the best of our knowledge, this is the first paper which takes prior information about *PoI* s into consideration and maximizes the utility of the geo-location perturbation mechanism while providing $\rho$ level of privacy to the user.

# 8 Acknowledgement

We would like to thank Dr. Aleksandra Korolova for being the guiding light throughout the course of this paper.

# A  Appendix

The domain D and range R is the *x*-axis discretized with step *δ*. Let *p* be the maximum value that should occur at the original location *i* = (0, 0). The probability values for output points *z*
at points $\epsilon$ (*δ*, ∞) are smaller than *p* but greater than points $\epsilon$ (−*δ*, −∞).

| i\z | -∞ | ... | −δ | 0 | δ | ... | +∞ |
|---|---|---|---|---|---|---|---|
| -∞ |  |  |  |  |  |  |  |
| : |  |  |  |  |  |  |  |
| -δ |  |  | p |  |  |  |  |
| 0 |  |  | ↓ | p | ↑ |  |  |
| δ |  |  |  |  | p |  |  |
| : |  |  |  |  |  |  |  |
| ∞ |  |  |  |  |  |  |  |

Now using the privacy constraint–

$\sum_{z=-\infty}^{\infty} P(i, z, \psi) = 1$

$\sum_{z=-\infty}^{-\delta} P(i, z, \psi) + p + \sum_{z=\delta}^{\infty} P(i, z, \psi) = 1$  ... (1)

or $A + B + C = 1$

$A = \sum_{z=-\infty}^{-\delta} P(i, z, \psi); B = p; C = \sum_{z=\delta}^{\infty} P(i, z, \psi)$

For $C$, we can use differential privacy constraint-

$$\frac{P(i,K(i)=z,\psi)}{P(j,K(j)=z,\psi)} \leq e^{\rho}; |i-j| \leq \delta$$

$i = (0,0), P(0,0,\psi) = p$ and $j = (\delta, 0)$ so we can write $P(\delta, z, \psi)$–

$$P(\delta, z, \psi) \leq p.e^{-\rho}$$

For $P(2\delta, z, \psi)$, we have–

$P(2\delta, z, \psi) \leq p.e^{-2\rho}$ and in general,

$P(x\delta, z, \psi) \leq p.e^{-x\rho}$, therefore we can rewrite $C$ in eq. (1) as

$\sum_{x=\delta}^{\infty} p.e^{x\rho}$     ... (2)

For part $A$ of eq. (1), we have $P(0, -\delta, \psi) < P(0, \delta, \psi) < p$. With utility constraint of $min.|z-i|$, along with the constraint of having higher probability of outputting points in the direction of prior, we can say that after some point $\alpha\delta$ it would be better to output points near the original location $i$ either in the direction opposite to the prior, i.e.,

$P(0, -\delta, \psi) \geq P(0, \alpha\delta, \psi) = p.e^{-\alpha\rho}$.

While maintaining the differential privacy constraint for the points $-\delta, -2\delta, ...$, we can write–

$$P(0, -\delta, \psi) \geq e^{-\rho}.P(0, -2\delta, \psi), \text{ or}$$

$p.e^{-\alpha\rho}.e^{\rho} \geq P(0, -2\delta, \psi)$ and in general–

$e^{(x-\alpha)\rho}.p \geq P(0, -(x-1)\delta, \psi)$, therefore we can write $A$ in eq. (1) as

$\sum_{x=-\infty}^{-\delta} e^{(x-\alpha)\rho}.p$     ... (3)

Combining $(1), (2)$ and $(3)$-

$\sum_{x=-\infty}^{-\delta} e^{(x-\alpha)\rho}.p + p + \sum_{x=\delta}^{\infty} p.e^{x\rho} \leq 1$

Solving this with $\delta = 1$ we get,

$p \leq \frac{(1-e^{-\rho})}{1+e^{-(\alpha+1)\rho}}$

# B  Appendix

For query 2, using the constraints we can write–

$\sum_{z=-\infty}^{\infty} P(i, z, \psi) = 1$

$$\sum_{z=-\infty}^{-\delta} P(i,z,\psi) + p + \sum_{z=\delta}^{L} P(i,z,\psi) + \sum_{z=L}^{2L} P(i,z,\psi) + p + \sum_{z=2L}^{\infty} P(i,z,\psi) = 1 \quad \ldots (1)$$

or $A + B + C + D + E + F = 1$

Since we are interested in the magnitude of the probability, for the sake of simplicity, we can safely apply same approximation before $i$ and after $2L$, and using the symmetry around $L$, we can write–

$$p \leq \frac{1}{\frac{e^{-\alpha\rho}+e^{-2\alpha.L.\rho}}{1-e^{-\alpha\rho}} + 2 + \frac{2.e^{-\rho}[1-(e^{-\rho.L})]}{1-e^{-\rho}}}$$

or approximately– $p \leq \frac{(1-e^{-\rho})}{2(1+e^{-(\alpha+1)\rho})}$, when $\delta = 1$